\documentclass[superscriptaddress,amsfonts]{revtex4}
\usepackage{epsf}

\newcommand{\bd     }{\begin{displaymath}}
\newcommand{\ed     }{\end{displaymath}}

\newcommand{\bra    }{\langle}
\newcommand{\ket    }{\rangle}

\newcommand{\atanh  }{{\rm{atanh}}}

\newcommand{\bh     }{\mbox{\boldmath$h$}}

\newcommand{\bzero  }{\mbox{\boldmath$0$}}

\newcommand{\bx     }{\mbox{\boldmath$x$}}
\newcommand{\bn     }{\mbox{\boldmath$n$}}
\newcommand{\by     }{\mbox{\boldmath$y$}}
\newcommand{\br     }{\mbox{\boldmath$r$}}
\newcommand{\bt     }{\mbox{\boldmath$t$}}

\newcommand{\cI     }{{\cal I}}
\newcommand{\cM     }{{\cal M}}
\begin{document}
\title{\bf Magnetization enumerator of real-valued symmetric channels
in Gallager error-correcting codes}

\author{\bf N.S.~Skantzos, J.~van Mourik and D.~Saad}
\affiliation{Neural Computing Research Group
 Aston University, Birmingham, B4 7ET, UK}

%%*******1*********2*********3*********4*********5*********6*********7********%%
\begin{abstract}
%%*******1*********2*********3*********4*********5*********6*********7********%%
Using the magnetization enumerator method, we evaluate the practical and
theoretical limitations of symmetric channels with real outputs.
Results are presented for several regular Gallager code constructions.
\end{abstract}

\maketitle
%%*******1*********2*********3*********4*********5*********6*********7********%%
%% Intro                                                                      %%
%%*******1*********2*********3*********4*********5*********6*********7********%%
Error correcting codes play a central role in modern communication,
especially in noisy media such as in satellite and mobile
communication. A broad range of error correcting codes are on offer;
they vary significantly in their practical and theoretical performance
depending on the specific code chosen within a given code ensemble.
Evaluating the limitations of specific code constructions is important
for determining the code efficiency and for optimizing channel
performance. One of the leading code ensembles to date is the family
of Low Density Parity Check Codes~\cite{mackay,richardson} (LDPC),
which attracted significant interest both within and outside of the
information theory community.

Methods of statistical physics have been recently employed to study
the typical performance of various coding schemes, most notably of Low
Density Parity Check
Codes~\cite{kabashima,murayama,vanmourik,montanari2,franz,tanaka}.  Such
studies have led to precise estimations of critical channel-noise
levels (beyond which decoding is not possible) and also provided
additional insight through the physical interpretation of various
decoding schemes. The emerging picture for Gallager-type codes is that
for sufficiently small noise levels decoding is possible and the
error-free (ferromagnetic) state is the only solution.  For higher
noise levels one finds a transition to a regime where suboptimal
solutions are created (spinodal or dynamical transition) and where
existing practical decoding algorithms fail to find the most probable
solution.  For higher noise levels, a second transition occurs
(thermodynamic transition) where the error-free solution ceases to be
dominant.  This marks the upper theoretical bound for error-free
communication using the code specified. The number of equally plausible solutions to the
decoding problem thereafter is exponential in the number of degrees of
freedom.  The thermodynamic transition approaches Shannon's limit 
with an increasing number of parity checks per bit.

%%*******1*********2*********3*********4*********5*********6*********7********%%
%% Technical                                                                  %%
%%*******1*********2*********3*********4*********5*********6*********7********%%
One of the most important aspects in the decoding problem is the
enumeration of possible solutions, as it provides a direct indication
to the practical and theoretical performance of various decoding
methods.  A method for carrying out the analysis have been reported
recently in~\cite{vanmourik}; the new approach generated interest in
the application of the same method to other channel types
characterized by real noise and real output values studied in this
report. In particular we study the cases of the Gaussian and Laplace
channels which are of high practical relevance, and are used as
standard benchmark channels for evaluating the performance of codes.
We consider the channels to be symmetric, i.e., the probabilities
$P(z_{\rm out}|z_{\rm in})=P(-z_{\rm out}|-z_{\rm in})$, where $z_{\rm
  out}$ and $z_{\rm in}$ are the input and output channel values
respectively.  We also consider information vectors representing
binary messages $\bt\in\{0,1\}^N$ to be encoded via Gallager's scheme.

A Gallager code is defined by the binary parity check matrix $A=[C_1|C_2]$ of
dimensionality $(M-N)\times M$, which is a concatenation of two submatrices.
A {\em regular} $(K,C)$ Gallager code has a fixed number $K$ of non-zero
elements per row in $A$, and a fixed number $C$ of non-zero elements
per column. It follows that $C\equiv K(M-N)/M$.

The message vector $\bt\in\{0,1\}^N$ is encoded to the codeword
$\bx=G^T\bt\in\{0,1\}^M$ prior to transmission, using the generator
matrix $G=\left[ I | (C_2^{-1}C_1)^T\right]$. This construction ensures
that $AG^T=0$ (mod 2). Redundancy in the codeword, in the case of
unbiased messages, is measured by the rate $R\equiv N/M=1-C/K$.  After
transmission of the codeword through the noisy channel, the following
message is received
\begin{equation}
\by=\bx+\bn^r
\label{eq:real_output}
\end{equation}
where $\bn^r\in I\!\!R^M$ represents the real channel noise which, in
the case of Gaussian and Laplace channels, has the distribution:
\begin{equation}
{\rm Gaussian:}
\hspace{10mm}
P(n^r_i)=\frac{1}{\sqrt{2\pi \sigma^2}}\
\exp[-\frac{1}{2\sigma^2}(n^r_i)^2]
\label{eq:gaussian_prob_noise}
\end{equation}
\begin{equation}
{\rm Laplace:}
\hspace{10mm}
P(n^r_i)=\frac{1}{2\lambda}\ \exp[-\frac{|n_i^r|}{\lambda}] \ ,
\label{eq:laplace_prob_noise}
\end{equation}
per bit $i$. The channel noise level is measured by the parameters
$\sigma$ and $\lambda$ respectively.

Note that since the channel noise $\{n_i^r\}$ consists of real-valued
variables (unlike the binary symmetric channel studied
in~\cite{vanmourik}), the channel output $\by$ is also real-valued and
the evaluation of the syndrome vector cannot be based on simple modulo
2 operations. Decoding is carried out by applying Bayes rule and the 
corresponding noise model to calculate $P(x=\pm1|y)$ for each bit.
To bring the problem back to a binary setting, we follow the
procedure of~\cite{mackay}, and
consider a fictitious channel where the sent message $\bx$ is
corrupted by binary noise $\bn^f\in\{0,1\}^M$:
\begin{equation}
\br=\bx+\bn^f
\hspace{10mm} {\rm (mod\ 2)}
\end{equation}
Without loss of generality, one can take $\bn^f=\bx$ such that $\br=\bzero$ and
$P(\bn^r)=P(\by|\bx)=P(\by|\bn^f)$.

We denote the set of (fictitious) noise vectors $\bn$ that satisfy the
parity check equations $A\bn=A\br=\bzero$ by ${\cI}_{pc}=\{\bn|A\bn=\bzero\}$,
the parity check set. To infer the original message one needs to find the
original fictitious noise $\bn^f$ from the parity check set on the basis of its
statistics. The conditional probabilities of the fictitious noise variables
(that satisfy the parity checks) follow from applying Bayes rule (per bit $i$)
\begin{equation}
P(n_i|y_i)=\frac{P(y_i|n_i)\,P(n_i)}
  {\sum_{n_i^\prime}P(y_i|n_i^\prime)\,P(n_i^\prime)} \ .
\label{eq:Pny}
\end{equation}

It was shown (e.g., in~\cite{kabashima,murayama}) that this problem
can be cast into a statistical mechanical formulation replacing the
field $(\{0,1\},+{\rm mod~2})$ by the field $(\{1,-1\},\times)$ and by
suitably adapting the parity checks.
Using the fact that $n_i$ are Ising variables with prior $P(n_i)=1/2$, 
(\ref{eq:Pny}) can now be written as $P(\bn|\by)=\exp(\beta H(\bn))$ where the
{\em energy} $H(\bn)$ is up to a constant given by
\begin{equation}
H(\bn)=\sum_i \ln P(n_i|y_i)=\frac1d \sum_i n_i h_i,\hspace*{1cm}
h_i\equiv \frac{d}{2}\sum_{\tau\in\{-1,1\}}\tau\ln P(y_i|\tau)
\label{eq:hi}
\end{equation}
and $\beta=1$ (which corresponds to Nishimori's
condition~\cite{Nishi}).  To unify notation for the Gaussian and the
Laplace channel, we denote the channel degradation parameter variable
by $d$, where $d=\sigma^2,~\lambda$ for the Gaussian and Laplace
channel respectively.  With these definitions, for any symmetric
channel with real outputs, the local energies are staggered
magnetizations along the fields $h_i$, the distribution of which
follows from (\ref{eq:hi}) in combination with
(\ref{eq:gaussian_prob_noise}) and (\ref{eq:laplace_prob_noise})
respectively:
\begin{equation}
{\rm Gaussian:}
\hspace{10mm}
p(h_i)=\sqrt{\frac{\sigma^2}{2\pi}}\
e^{-\frac{\sigma^2}{2}(h_i-\frac{1}{\sigma^2})^2}
\label{eq:fields_G}
\end{equation}
\begin{equation}
{\rm Laplace:}
\hspace{10mm}
p(h_i)=\frac12~\delta(h_i-\lambda^{-1})+\frac{e^{-\frac{2}{\lambda}}}{2}~
\delta(h_i+\lambda^{-1})
+\Theta[\lambda^{-1}-|h_i|]~\frac12 e^{h_i-\lambda^{-1}}
\label{eq:fields_L}
\end{equation}
where $\Theta[x]$ is the Heavyside function returning 1 if $x\geq0$ and 0 if
$x<0$.

The entropy of the solutions to the decoding problem with a given magnetization
$m(\bn;\bh)=\frac1M\sum_i n_i h_i=m$ is
\begin{equation}
\cM(m)=\frac1M\left\bra\ln\sum_{\bn\in\cI _{pc}(\bn,\bn^f;A)}~
\delta[m-m(\bn;\bh)]\right\ket \ .
\label{eq:S_def}
\end{equation}
Averages in~(\ref{eq:S_def}) are taken over the fields $\{h_i\}$
((\ref{eq:fields_G}) or (\ref{eq:fields_L})), the parity check constructions
$A$, whereas the original fictitious noise $\bn^f$ is gauged away using the
transformations $n_i\to n_in_i^f$ and $y_i\to y_i n_i^f$.
In order to perform the averages, we employ the replica identity:
%
%\begin{equation}
$
\bra \ln \cM(m)\ket=\lim_{n\to 0}\frac1n\ln\bra \cM^n(m)\ket \ .
$
%\end{equation}
%
In the limit $n\to 0$ and within the replica symmetric assumption
(shown to be exact in this case for obtaining the theoretical critical
noise levels~\cite{vanmourik}; for technical details also see
e.g.~\cite{kabashima,wongsherrington}), we find that
\begin{equation}
\cM(m)={\rm Extr}_{\pi,\hat{\pi},\hat{m}}\left\{-\hat{m}m-\frac{C}{K}\ln 2+
\frac{C}{K}I_1[\pi]-CI_2[\pi,\hat{\pi}]+I_3[\hat{\pi};\hat{m}]\right\}
\label{eq:S}
\end{equation}
with
\begin{equation}
I_1[\pi]=\int \prod_{k=1}^K\{dx_k \pi(x_k)\}~\ln[1+\prod_{k=1}^K x_k]
\label{eq:I1}
\end{equation}
\begin{equation}
I_2[\pi,\hat{\pi}]=\int dxd\hat{x}~\pi(x)\hat{\pi}(\hat{x})\
\ln(1+x\hat{x})
\label{eq:I2}
\end{equation}
\begin{equation}
I_3[\hat{\pi};\hat{m}]=\int \prod_{c=1}^C \{d\hat{x}_c \hat{\pi}(\hat{x}_c)\}\
\left\bra\ln\sum_{\tau=\pm} e^{\tau\hat{m}
h}\prod_{c=1}^C(1+\tau \hat{x}_c)\right\ket_{h}
\label{eq:I3}
\end{equation}
The functional extremization problem in (\ref{eq:S}) results in the following
saddle point equations
\begin{eqnarray}
\hat{\pi}(\hat{x})&=&\int \prod_{k=1}^{K-1}\{dx_k \pi(x_k)\}\
\delta[x-\prod_{k=1}^{K-1} x_k] 
\label{eq:pi_hat} \\
\pi(x)&=&\int \prod_{c=1}^{C-1}\{d\hat{x}_c \hat{\pi}(\hat{x}_c)\}~\left\bra \delta\left[x-
\tanh[\hat{m}h+\sum_{c=1}^{C-1}\atanh(\hat{x}_c)]\right]\right\ket_h
\label{eq:pi} \\
m&=&\int \prod_{c=1}^C\{ d\hat{x}_c~\hat{\pi}(\hat{x}_c)\}~
\left\bra
 \frac{h\sum_{\tau=\pm} \tau e^{\hat{m}\tau h}
       \prod_{c} (1+\tau \hat{x}_c)}
      {\sum_{\tau=\pm} e^{\tau\hat{m} h}
       \prod_{c} (1+\tau \hat{x}_c)}\right\ket_h
\label{eq:m}
\end{eqnarray}
Equations (\ref{eq:pi_hat}) and (\ref{eq:pi}) are the infinite system
equivalent of the so-called density-evolution equations
\cite{richardson}, and iteratively converge to the stationary
distributions $\pi^*(x)$, $\hat{\pi}^*(x)$.
Equations~(\ref{eq:S},\ref{eq:m}) are then evaluated for these
stationary distributions. One should note that the solution
$\hat{\pi}^*(\hat{x})= \delta[\hat{x}-1]$ and $\pi^*(x)=\delta[x-1]$
always exists, and has a magnetization $m_0=\bra h\ket_h$ (\ref{eq:m})
and zero entropy. This (ferromagnetic) solution corresponds to perfect
retrieval ($\bn^f$ after the gauge), and should be compared to
alternative solutions, if they exist, which correspond to the other
(sub-optimal) noise candidates in $\cI_{pc}(\bn,A)$.

In the limit $K,C\!\to\!\infty$ (while keeping the rate $R$ finite) it
is possible to derive these alternative solutions for all values of
$m$, analytically.  This case is, however, of little practical
interest, and will not be discussed in this paper.

For finite $K,C$, alternative analytic solutions can no longer be
obtained for both Gaussian and Laplace channels, and one has to solve
the saddle point equations numerically to obtain $\cM(m)$.

The critical noise values of the thermodynamic and spinodal
transitions now follow directly from the graphs of $\cM(m)$, for
different values of $d$.

\begin{figure}[th]
%\vspace{-25mm}
\setlength{\unitlength}{1.4mm}
\begin{picture}(120,55)
\put( 30,  0){\epsfysize=45\unitlength\epsfbox{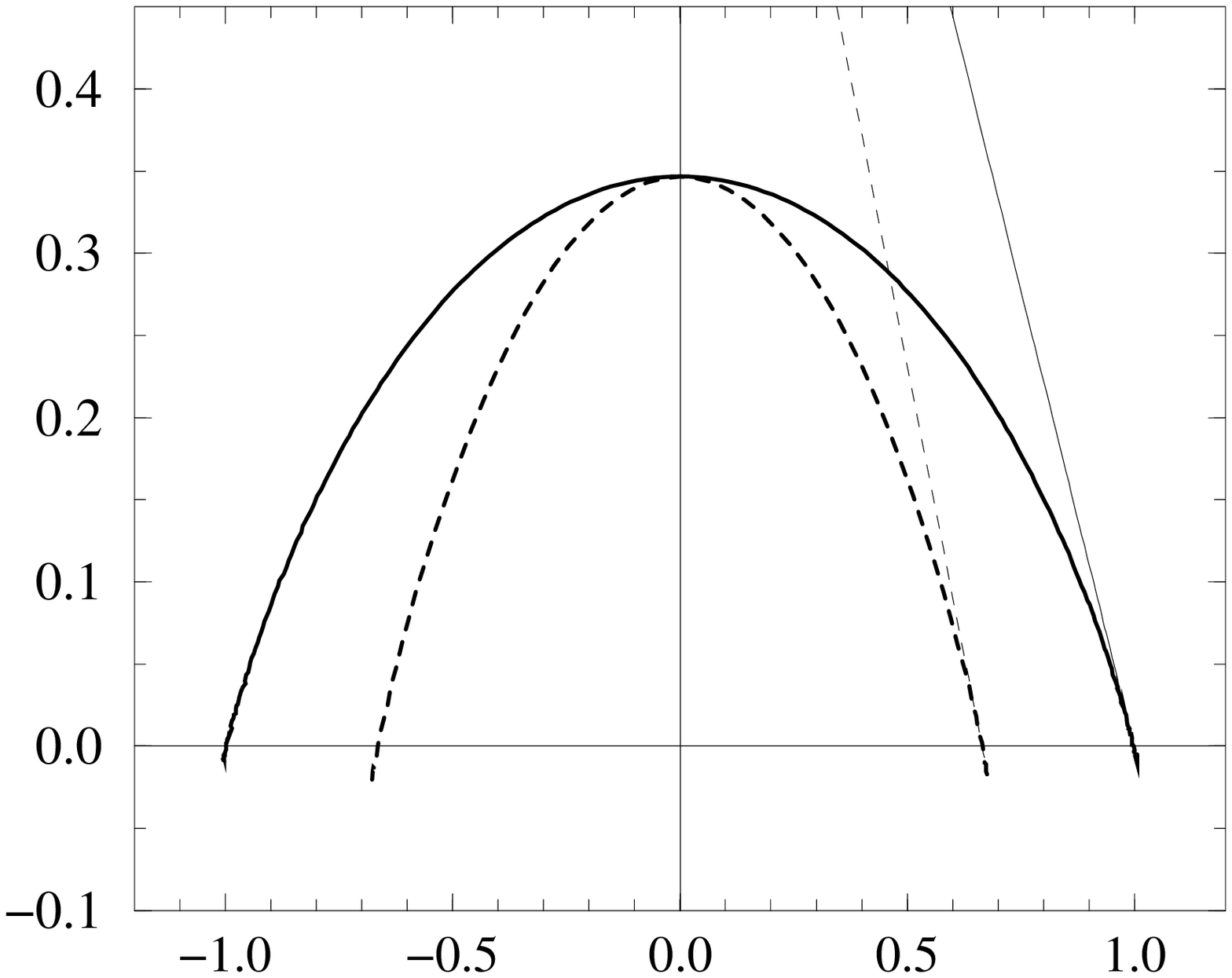}}
\put(25,25){$\cM(m)$}
\put(60,2){$m$}
\end{picture}
\caption{\small
  The magnetization enumerator for the Gaussian (solid curve) and the
  Laplace (dashed curve) channels for a $(K,C)=(6,3)$ code at the
  thermodynamic transition noise levels ($\sigma^2_c=0.899$ ,
  $\lambda_c=0.712$). For both channels the maximum number of
  solutions is $\cM(0)=R\ln 2$.  The energy of solutions is given by
  $E(m)=-\frac1d m$, while their free energy at the Nishimori
  temperature~\cite{Nishi} is up to a constant given by the orthogonal
  distance to the straight lines. At the thermodynamic transition
  point, these are tangents to $\cM(m)$ at $m_\star=m_0$(=1 and 0.665
  for the two channels respectively). }
\label{fig:figure}
\end{figure}

As explained in full detail in~\cite{vanmourik}, the theoretical
critical degradation value $d=d_{c}$ is reached, for Maximum A
Posteriori (MAP) and typical set decoding methods, when the
magnetization at which the entropy of the sub-optimal solutions
vanishes, coincides with that of the ferromagnetic solution. For
finite temperature decoding at the Nishimori temperature~\cite{Nishi},
the critical noise value coincides with the thermodynamic transition
at which the ferromagnetic solution ceases to be dominant, and
$\cM(m)$ has a slope $\left.{\partial\cM(m)/\partial m}\right|_{m=\bra
  h\ket_h}=-1/d$ at $m=m_0=\bra h\ket_h$ (see Fig.~\ref{fig:figure}).

Of more practical interest is the limiting practical noise level
$d=d_{d}$, above which practical algorithms such as density
evolution~\cite{richardson} break down. This transition is signaled by
the emergence of suboptimal solutions for
eqns.~(\ref{eq:pi_hat},\ref{eq:pi}). These correspond to local minima
of the free energy in which the algorithm gets trapped; this is known
as a spinodal point or dynamical transition. The noise level
$d=d_{d}$ can be obtained from $\cM(m)$; it is the smallest positive
value $d$ for which there exists a value $m^*$ such that
$\left.\partial \cM(m)/\partial m\right|_{m^*}=-1/d$. This
typically happens at a value of $m^*$ for which $\cM(m^*)<0$,
such that the practical transition point is upper-bounded by the
thermodynamic transition point, which is in turn upper-bounded by
Shannon's information-theoretic limit, and the following inequalities
hold: $d_{d}\leq d_c\leq d_{S}$.

In Fig.\@ \ref{fig:figure} we present the magnetization enumerator 
for the Gaussian and Laplace channels for a $(K,C)=(6,3)$ code 
at the thermodynamic transition. 
The maximum number of solutions to both channels depends only on the 
code rate $R$ as $\cM(0)=R\ln 2$.  %

It should be noted that values of $\cM(m)<0$ are unphysical, and are
an artifact of the replica symmetric assumption. Nevertheless, this
region turns out to be relevant for the determination of the dynamical
(spinodal) transition.  This can be understood by the fact that the
replica symmetric fixed point equations are the exact infinite size
equivalent of the practical density evolution
equations~\cite{richardson}. Therefore, although quantities related to
the $\cM(m)<0$ region, are unphysical and should be corrected by a
refined replica symmetry ansatz, the region $\cM(m)<0$ and the fixed
point equations associated with it, still allow us to determine the
dynamical transition point.

The calculated critical noise levels are presented in the
Table~\ref{tab:critical_values} for several code constructions and for
the Gaussian and Laplace channels. These values are in excellent
agreement with those obtained independently in~\cite{tanaka} using a
different method, as well as with the corresponding practical
upper-bounds of~\cite{richardson} obtained using density evolution. 
Note that we have also
presented the critical degradation values for different code
constructions of identical rates, to illustrate the opposite
tendencies for the theoretical and practical critical values with
increasing $K$ and $C$.  Increasing $K$ and $C$ (keeping $K/C$ fixed)
pushes the thermodynamic critical value closer to Shannon's
information-theoretic limit, but adversely affects the practically
admissible degradation value. This is in agreement with the common
belief that code constructions with higher connectivity are less
practical.
%

%\vspace{10mm}
\begin{table}
\hspace{0mm}
\begin{tabular}{||c|c|c|c|c||} \hline\hline
  $(K,C)$ & $R$    & $\sigma^2_{d}$ &  $\sigma^2_{c}$ & $\sigma^2_{S}$\\
  \hline \hline
  (6,3) & 0.5   & 0.775 & 0.899 & 0.958 \\ \hline
  (5,3) & 0.4   & 1.017 & 1.253 & 1.321 \\ \hline
  (6,4) & 0.333 & 1.020 & 1.666 & 1.681 \\
  (9,6) & 0.333 & 0.379 & 1.679 & 1.681 \\ \hline 
  (4,3) & 0.25  & 1.598 & 2.325 & 2.401 \\
  (8,6) & 0.25  & 0.880 & 2.396 & 2.401 \\ \hline
\end{tabular}
%\hspace{20mm}
\begin{tabular}{||c|c|c|c|c||} \hline\hline
  $(K,C)$ & $R$    & $\lambda_{d}$ &  $\lambda_{c}$ & $\lambda_{S}$\\
  \hline \hline
  (6,3) & 0.5   & 0.651 & 0.712 & 0.752 \\ \hline
  (5,3) & 0.4   & 0.773 & 0.875 & 0.914 \\ \hline
  (6,4) & 0.333 & 0.782 & 1.045 & 1.055 \\
  (9,6) & 0.333 & 0.661 & 1.048 & 1.055 \\ \hline
  (4,3) & 0.25  & 1.018 & 1.260 & 1.298 \\
  (8,6) & 0.25  & 0.619 & 1.271 & 1.298 \\ \hline
\end{tabular}
\caption{Values of the critical noise levels
(spinodal and thermodynamic transitions) of the Gaussian and Laplace
channels for various regular $(K,C)$ Gallager codes. For comparison, 
in the last column we present Shannon's information-theoretic bound.}
\label{tab:critical_values}
\end{table}

In this report we have shown how the magnetization enumerator
formalism~\cite{vanmourik} can be easily extended to real-valued
channels, in order to obtain both theoretical and practical critical
values for the degradation parameter.  Following the method
presented in~\cite{mackay} we have mapped the real valued channel onto
an equivalent fictitious binary channel, and employed methods of
statistical physics to calculate the magnetization enumerator for the
Gallager code ensemble.  The magnetization enumerator is instructive
in the way it nicely links the various decoding
schemes~\cite{vanmourik} and facilitates the derivation of both
practical and theoretical critical noise levels.

Using Nishimori's gauge theory, the theoretical critical noise levels
can be shown to be exact, while our practical critical degradation
parameters, are in excellent agreement with those obtained known in
the literature~\cite{tanaka,richardson}, when available.

Studying the magnetization enumerator further, beyond the
practical limiting noise level may provide additional insight into the
decoding complexity  the performance of Gallager type codes.

\begin{acknowledgments}
  We would like to thank Toshiyuki Tanaka for helpful
  suggestions. Support from EPSRC research
  grant GR/N63178 is acknowledged.
\end{acknowledgments}

\end{document}